%% file: rbb_brl.tex
\documentclass[12pt,a4paper,dvips]{article}
\usepackage{a4p}
\usepackage{cite,mcite}
\usepackage{graphicx}
\usepackage{physics}
\usepackage{l3_title,ifthen,Lep}
\journalname{European Physics Journal}
\preprint{99-121}
\date{August 27, 1999}
\Lep{1}

\begin{document}
 
%
%
\def\Brbcbarl 
{\ensuremath{\Br({\mathrm{b}}\ra{\mathrm{\overline{c}}}\ell\nu{\mathrm{X}}) }}
\def\blnuX  {\ensuremath{\mathrm{b} \ra \ell\nu\mathrm{X}}}
\def\btnuX  {\ensuremath{\mathrm{b} \ra \tau\nu\mathrm{X}}}
\def\bccs  {\ensuremath{\mathrm{b} \ra \mathrm{c}\mathrm{\overline{c}}\mathrm{s}}}
\def\Brbcl  {\ensuremath{\Br(\mathrm{b} \ra \mathrm{c} \ra \ell) }}
\def\BrBl  {\ensuremath{\Br(\mathrm{B} \ra \ell\nu\mathrm{X}) }}
\def\BrLl  {\ensuremath{\Br(\mathrm{\Lb} \ra \ell\nu\mathrm{X}) }}
\def\Brbcbarl  {\ensuremath{\Br(\mathrm{b} \ra \mathrm{\overline{c}} \ra \ell) }}
\def\Brbjl  {\ensuremath{\Br(\mathrm{b} \ra \Jpsi \ra \ell) }}
\def\Brbtaul  {\ensuremath{\Br(\mathrm{b} \ra \tau \ra \ell) }}
\def\taub  {\ensuremath{\tau_\mathrm{b}}}
\def\cb  {\ensuremath{c_\mathrm{b}}}
\def\cc  {\ensuremath{c_\mathrm{c}}}
\def\cuds  {\ensuremath{c_\mathrm{uds}}}
\def\epsb  {\ensuremath{\epsilon_\mathrm{b}}}
\def\epsc  {\ensuremath{\epsilon_\mathrm{c}}}
\def\epsuds  {\ensuremath{\epsilon_\mathrm{uds}}}
\def\Rb  {\ensuremath{R_\mathrm{b}}}
\def\Rc  {\ensuremath{R_\mathrm{c}}}
\def\Nc  {\ensuremath{N_\mathrm{c}}}
\def\mc  {\ensuremath{m_\mathrm{c}}}
\def\msp {\ensuremath{m_\mathrm{sp}}}
\def\pf  {\ensuremath{p_f}}
\newcommand{\Do}  {\ensuremath{\mathrm{D^0}}}
\newcommand{\Ko}  {\ensuremath{\mathrm{K^0}}}
\newcommand{\Kos}  {\ensuremath{\mathrm{K^0_S}}}
\newcommand{\Dp}  {\ensuremath{\mathrm{D^+}}}
\newcommand{\Ds}  {\ensuremath{\mathrm{D_s}}}
\newcommand{\Lc}  {\ensuremath{\mathrm{\Lambda_c}}}
\def\stasys{\mbox{$\;$(stat.+sys.)}}

\begin{titlepage}

\title{\boldmath Measurement of $\Rb$ and $\Brbl$ at LEP Using 
Double-Tag Methods}

\author{The L3 Collaboration}

\begin{abstract}

  We present a combined measurement of 
$\Rb = \Gamma(\mathrm{Z \rightarrow b\bar{b}}) / 
\Gamma(\mathrm{Z} \rightarrow \mbox{hadrons})$ 
and the semileptonic branching ratio of
b quarks in Z decays, $\Brbl$, using double-tag methods. Two analyses are 
performed 
on one million hadronic Z decays collected in 1994 and 1995. The first analysis 
exploits the capabilities of the silicon microvertex detector. The tagging 
of b-events is based on the large impact parameter of tracks from weak b-decays with
respect to the $\ee$ collision point.  In the second analysis, a
high-$\pt$ lepton tag is used to enhance the b-component in the sample
and its momentum spectrum is used to constrain the model dependent 
uncertainties in the semileptonic b-decay. The analyses are combined in order 
to provide precise determinations of $\Rb$ and $\Brbl$:

    $$\Rb = 0.2174 \pm 0.0015\stat \pm 0.0028\sys;$$
    $$\Brbl = (10.16 \pm 0.13\stat \pm 0.30\sys)\%.$$
  
\end{abstract}

\submitted

\end{titlepage}
 
\section{Introduction}
The Z partial width into b-quark pairs is a relevant parameter
for precision studies of the Standard Model (SM) \cite{standard_model}.
Due to the high mass of the top quark and the large 
top-bottom weak coupling, the process $\Zo \ra \bbbar$ receives sizable 
contributions from vertex corrections in the SM. The partial width is usually 
measured as its relative contribution to the $\Zo$ hadronic width, 
$\Rb = \Gamma(\mathrm{Z \rightarrow b\bar{b}}) /
\Gamma(\mathrm{Z} \rightarrow \mbox{hadrons})$, since many experimental and
theoretical uncertainties cancel when forming this ratio. A deviation from the 
predicted SM value $0.2158 \pm 0.0002$ \cite{zfitter} would point to the 
existence of additional vertex corrections and therefore would be a signal for 
new physics.

 The semileptonic branching ratio of b-hadrons, $\Brbl$,
can be expressed as:
\begin{eqnarray}
   \Brbl = \frac {\Gamma_{\blnuX}}{\Gamma_{\rm all} } =
   \frac {\Gamma_{\blnuX}}
        {2\Gamma_{\blnuX} + \Gamma_{\btnuX} + \Gamma_{\rm had}},
\end{eqnarray}

\noindent
where $\Gamma_{\blnuX}$ is the semileptonic decay width, $\ell$ being an 
electron or a muon, $\Gamma_{\btnuX}$ is the semileptonic component 
involving tau production and $\Gamma_{\rm had}$ is the partial width from 
purely hadronic decays. Present measurements of $\Brbl$ at LEP \cite{lepew99} 
are in slight disagreement with model-independent measurements performed at 
the $\Upsilon(4S)$ \cite{4S_semilep}. Z decays into b-quarks allow for the
presence of b-baryons, which have a lower semileptonic 
branching fraction \cite{lambdab_semilept}. Measurements of the average number 
of charmed hadrons in b-decays, $\Nc$, provide an indirect estimate of 
the $\bccs$ component. The measured values in the $\Nc-\Brbl$ plane show that a 
discrepancy is still present between the $\Upsilon(4S)$ and LEP results, 
as well as between the $\Upsilon(4S)$ results and theoretical predictions 
\cite{neubert}.

In this paper, we present a combined measurement of $\Rb$ and $\Brbl$ obtained 
with the L3 detector \cite{bib-L3exp} at LEP. Features that distinguish the 
production of $\mathrm{b\bar{b}}$ pairs from lighter quark production are: 
the long lifetime and hard fragmentation of b-flavoured hadrons, the large 
lepton momentum in semileptonic decays and the broad event shape caused by the 
large b-quark mass.
The b-tagging method using lifetime information relies on data taken with the 
L3 Silicon Microvertex Detector (SMD) \cite{smd} in 1994-1995.
The method using the characteristic semileptonic decays of b-hadrons
relies on the good lepton identification and lepton energy resolution of the
L3 detector. It requires lepton candidates with high momentum along and 
transverse to the direction of the associated jet, caused by the hard 
fragmentation and high mass of the decaying hadron.

\section{Hadronic Event Selection}\label{L3Det}

   Hadronic events are selected with criteria similar to the 
ones used for the measurement of the total hadronic cross section 
 \cite{bib-hadsell3}.
The basic requirements used there select 99.15\% of all
hadronic Z decays with a background of 0.15\% from other processes. A few
additional requirements ensure a good performance of the b-tagging techniques
on the events. These are:
\begin{itemize}
\item The number of reconstructed tracks must be larger than four.
\item The polar angle of the thrust axis, $\theta_T$,
reconstructed from calorimetric information, must be within a fiducial region
delimited by the barrel part of the detector, $|\cos{\theta_T}| < 0.7$.
\end{itemize}
With these additional criteria, 970k hadronic events are selected from data,
corresponding to an integrated luminosity of 71 pb$^{-1}$.

Hadronic Z decays are simulated using the JETSET \cite{bib-jetset73} generator 
and passed through a simulation of the L3 detector \cite{bib-geantl3}. The 
simulation takes into account the inefficiencies and resolutions 
of the different subdetectors as a function of time, weighted according to 
the integrated luminosity in data.

\section{Double-Tag Methods}

 An event is first split into two hemispheres defined by
the plane normal to the thrust axis. Separately for each hemisphere, a set of 
criteria is applied in order to significantly enhance the purity of b-events.
A hemisphere satisfying those criteria is declared to be ``tagged''.
The number of tagged hemispheres,  $N_t$, is related to the total
number of hadronic events, $N_{had}$, by the following equation:

\begin{equation}
\frac{N_t}{2 N_{had}} = \tilde{\Rb} \epsb + \tilde{\Rc} \epsc 
+ \left( 1 - \tilde{\Rc} - \tilde{\Rb} \right) \epsuds \label{eq1},
\end{equation}

\noindent
where $\epsb$, $\epsc$ and $\epsuds$ are
the tagging efficiencies for b, c and light quark hemispheres.
The parameter $\tilde{\Rb}$ represents the ratio of cross sections 
$\sigma(\mathrm{b\bar{b}}) / \sigma(\mbox{hadrons})$. It is related to
$\Rb$ by $\Rb=\tilde{\Rb} + 0.0003$.
The correction takes into account the contribution from photon exchange
\cite{zfitter}.
The shift is also present for $\Rc$, but its effect on the measurement can be 
ignored for samples of reasonable b-purity. Therefore we will assume
$\tilde{\Rc}=\Rc$ in the following.
The number of events with both hemispheres tagged, $N_{tt}$, is given by:

\begin{equation}
\frac{N_{tt}}{N_{had}} = \cb \tilde{\Rb} \epsb^2 + \cc \Rc \epsc^2
+ \cuds \left( 1 - \Rc - \tilde{\Rb} \right) \epsuds^2 \label{eq2},
\end{equation}

\noindent
where the additional factors $\cb$, $\cc$ and $\cuds$, called hemisphere
correlation factors, quantify residual correlations between the two
hemispheres, which lead to a deviation from the simple power law
reduction of the efficiencies.

Two parameters can be determined directly from data using the two 
experimentally measured ratios: $N_t/(2 N_{had})$ and $N_{tt}/N_{had}$; we 
choose to extract the parameters $\Rb$ and $\epsb$.
The relative rate of c production, $\Rc$, 
is constrained to its experimental value $\Rc=0.1734 \pm 0.0048$ \cite{pdg98}. 
The efficiencies for quarks lighter than the b-quark, as well as the hemisphere 
correlation factors, are taken from the Monte Carlo (MC) simulation.
However, only the factor $\cb$ is relevant to the analysis for samples of good
b purity ($\cc = \cuds = 1$). Typical values of the MC parameters for 
high purity tags are $\epsc \approx 2\%$, $\epsuds \approx 0.5\%$ and 
$\cb \approx 1$.

   If a different b-tagging algorithm is applied, three additional equations 
can be derived:
\begin{eqnarray}
\frac{N_{t^\prime}}{2 N_{had}} & = & 
 \tilde{\Rb} \epsb^\prime + \Rc \epsc^\prime
+ \left( 1 - \Rc - \tilde{\Rb} \right) \epsuds^\prime \label{eq3}, \\
\frac{N_{{t^\prime}{t^\prime}}}{N_{had}} & = &  
\cb^\prime \tilde{\Rb} \epsb^{\prime~2} + \Rc \epsc^{\prime~2}
+ \left( 1 - \Rc - \tilde{\Rb} \right) \epsuds^{\prime~2} \label{eq4}, \\
\frac{N_{{t}{t^\prime}}}{2 N_{had}} & = &
\cb^{\prime\prime} \tilde{\Rb} \epsb \epsb^\prime
+ \Rc \epsc \epsc^\prime
+ \left( 1 - \Rc - \tilde{\Rb} \right) \epsuds \epsuds^\prime \label{eq5}.
\end{eqnarray}

\noindent
where the new efficiencies and correlations have similar meanings to the 
efficiencies and correlations defined for the tag $t$.
   The measurement can be performed by a global fit in which the values
of the five ratios: $N_t/(2~N_{had})$, $N_{tt}/N_{had}$, $N_{t^\prime}/(2~N_{had})$, 
$N_{t{t^\prime}}/(2~N_{had})$, $N_{{t^\prime}{t^\prime}}/N_{had}$ are used
to determine the values of $\Rb$, $\epsb$ and $\epsb^\prime$, while
$\cb$, $\epsc$, $\epsuds$, $\cb^\prime$, $\epsc^\prime$, 
$\epsuds^\prime$, $\cb^{\prime\prime}$ are constrained to the values
obtained from the MC simulation with their statistical and 
systematic errors.

    The two tagging methods applied in this analysis are an impact parameter 
tag whose efficiency is denoted by $\epsilon$ and a leptonic tag with 
efficiency $\epsilon^\prime$.

\section{Impact Parameter Analysis}

\subsection{Track and Primary Vertex Reconstruction}\label{TraVert}
The inner tracker of L3 reconstructs particle trajectories from hits in the two layers
of double-sided silicon sensors of the SMD, up to 62 measurements in the
central tracking chamber and two measurements in the Z chamber. These 
measurements are combined to obtain
the five parameters characterising the trajectory, {\it i.e.} its curvature in 
the $r-\phi$ plane, its transverse distance of closest approach
(DCA) to the vertex, its azimuthal angle at the DCA, its polar angle $\theta$ and 
the Z coordinate at the DCA. Their
covariance matrix is determined from the estimated single-point resolution function.
The most important parameters for this analysis are the DCA and its error,
$\sigma_{DCA}$. Small biases in the DCA itself are removed by recalibrating the
mean DCA value as a function of the azimuthal angle of the track and as a
function of the track position inside a sector of the central tracking 
chamber \cite{b_life}. The DCA
error is recalibrated using tracks with a high probability of coming from the
primary vertex. For these, the width of the DCA distribution for tracks with
high momentum, where multiple scattering is negligible, is used to determine
a factor that multiplies the calculated DCA error from the track fit. The
factor is found to be close to one, in agreement with an analysis performed
using high momentum tracks from $\mathrm{e^+ e^- \rightarrow e^+ e^-}$,
$\mu^+ \mu^-$ and $\tau^+ \tau^-$ events. Typical values for $\sigma_{DCA}$ are 
$30~\mu$m and $100~\mu$m for tracks with and without SMD information, respectively.
In addition, the contribution from multiple
scattering, not included in the error calculated in the track fit, is
estimated from the dependence of the distribution width on 
transverse momentum. It is found that the additional multiple scattering error
is $110/(p_\perp \sqrt{\sin{\theta}})$ $\mu$m
for tracks with a hit in the inner SMD layer and
$200/(p_\perp \sqrt{\sin{\theta}})$ $\mu$m for tracks without such hits, with
$p_\perp$ measured in GeV.

The average position of the LEP luminous region inside L3 is reconstructed using
tracks collected in hadronic events. The position and its error are averaged
over 200 consecutive hadronic events, in order to follow drifts in the 
beam position.  The result, 
called the beam-spot position, is used as a constraint in the  reconstruction of the 
primary vertex in each event, weighted by the r.m.s.~width of the luminous region in the horizontal ($110~\mu$m) and vertical
($20~\mu$m) directions.

For the reconstruction of primary vertices, tracks are selected using the
following criteria:
\begin{itemize}
\item A track must consist of at least 20 hits in the central tracking chamber.
\item At least one hit in the inner layer of the SMD must be included
in the track fit.
\item The DCA to the primary vertex has to be less than 1 mm.
\item The significance of the DCA, defined by the ratio of the DCA and its
error, has to be less than five.
\item The transverse momentum of the track has to be greater than $150 \MeV$.
\end{itemize}
The procedure uses an iterative method which starts from
the beam-spot position as an initial estimate of
the primary vertex position. At each step of the iteration, the
vertex is calculated with all tracks selected for that step.
If the $\chi^2$ probability of the
vertex is less than 0.05, the track with the largest contribution
to the $\chi^2$ is removed and the vertex is recalculated with the
remaining tracks. This procedure is repeated until the
$\chi^2$ probability of the vertex is at least 0.05 or
only three tracks are left. At each step, the beam-spot position is used
as a constraint.
With this procedure, a primary vertex is reconstructed in
99.5\% of the selected events. This efficiency is found to be independent of the
quark flavour within one per mill, using MC simulation.

The uncertainty on the vertex position 
depends on the azimuthal angle of the event thrust axis
and on the number and quality of the tracks retained for its determination. 
The average uncertainty in the horizontal
direction is $42~\mu$m for light-quark events and $77~\mu$m for b-events.
The worse resolution for
b-events is due to the unavoidable inclusion of b-decay tracks in the
vertex determination. The vertical position of the primary 
vertex is known with high precision ($20~\mu$m) due to 
the small vertical width of the beam spot.

\subsection{Heavy Quark Tagging Using Impact Parameters}

 The information from all tracks in a hemisphere is combined
to form a discriminating variable, $D$, which describes the likelihood that all
tracks come from the primary vertex. The sensitive single-track quantity
used for constructing $D$ is the impact parameter, defined as the
absolute value of the track's DCA with a sign that is positive if the track
intersects the direction of the associated jet in the direction of the jet's
total momentum, negative if it intersects opposite to that direction. 
The angular resolution for the jet direction is 40 mrad.

Tracks retained for the determination of $D$ have to fulfil the following quality
criteria:
\begin{itemize}
\item The angle, $\theta_j$, between the track and its associated jet axis must 
      satisfy $\cos\theta_j > 0.7$.
\item The track should have at least 30 hits spanning over a distance of at 
      least 40 wires in the central tracking chamber.
\item The DCA to the primary vertex
      has to be less than 1.5 mm for tracks with SMD information. This cut is 
      increased to 3 mm for tracks without SMD hits.
\item The angular separation of the track from the anode and cathode planes
      of the central tracking chamber, where the resolution is worse, must be 
      more than 11 mrad.
\item If a track uses no hits from the SMD, at least 2 out of 8 hits from the
      inner portion of the central tracking chamber should be used in the 
      track fit.
\end{itemize}
The retained tracks are then grouped into different classes according to the 
pattern of the associated SMD hits. Each class corresponds to a
different resolution function for the impact parameter measurement. 
The repartition among classes obtained in data is compared to the MC
simulation. The proportions agree in absolute value to within a percent.

The significance, $s$, is defined as the ratio of the impact parameter to
its error. The total impact parameter error is composed of the
error from the track fit, the multiple scattering contribution and the
contribution of the primary vertex error, all determined according to the 
procedure described in Section~\ref{TraVert}.

The discriminant variable
is constructed on the basis of a resolution function, $R(x)$, which describes the
probability that a track which comes from the primary vertex is measured to have
an apparent impact parameter significance $x$. The probability, $P$, of finding
a significance greater than the measured one, $s$, is given by:

\begin{equation}
P(s) = \frac{\int_{s}^{\infty} R(x) dx}{\int_{-\infty}^\infty R(x) dx}.
\end{equation}

  The combined probability for the $n$ tracks in the hemisphere is
$\prod_{i=1}^{n} P(s_i)$. The probability, ${\cal P}(n)$, of measuring a value greater 
than $\prod_{i=1}^{n} P(s_i)$ is:

\begin{equation}
{\cal P}(n) = \prod_{i=1}^{n} P(s_i) \sum_{j=0}^{n-1} 
\frac{\left( -\ln \prod_{i=1}^{n} P(s_i) \right)^j}{j!}.
\end{equation}

 We define the discriminant variable as $D = - \log_{10} {\cal P}(n)$. 
A tagged hemisphere must have a value of $D$ above some minimum value.
This definition of the discriminant cut ensures that the amount of background from 
hemispheres which have tracks consistent with the primary vertex is $10^{-D}$, 
independent of $n$.

\begin{figure}[htbp]
\begin{center}
\includegraphics[width=\textwidth]{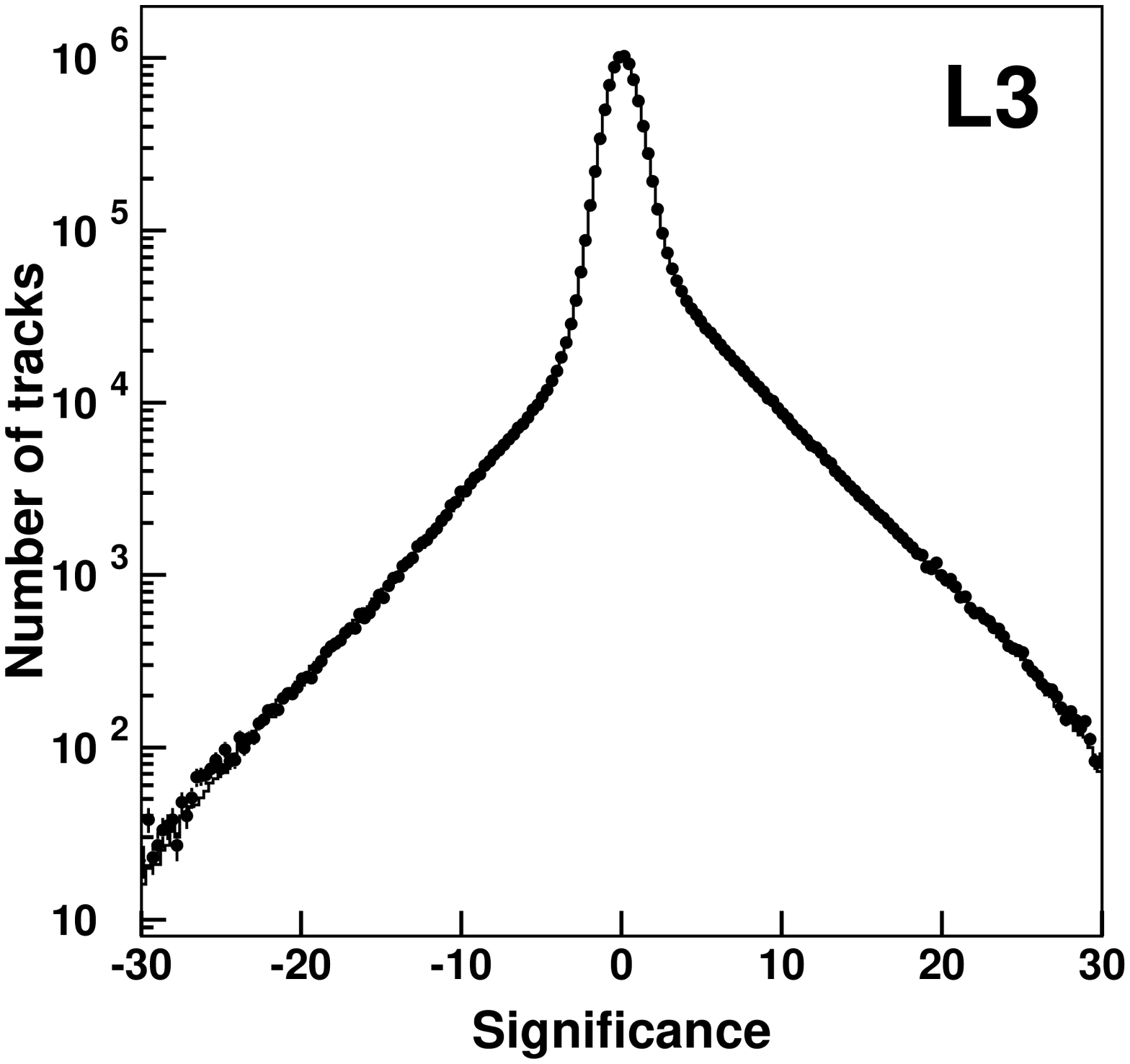}
\end{center}
\caption{\label{resf} Distribution of the significance.
  The points are the data and 
  the histogram is the MC prediction for a value of $\Rb=0.217$.}
\end{figure}

The resolution function $R$ is determined for
each class by a fit to the significance distribution of all tracks with negative 
significance in data. The 
r.m.s.~width of the significance distribution in each class is compatible with 1.0, 
but the distributions have substantial tails. Therefore, a model of the
resolution function is constructed as a sum of two Gaussian functions and an
exponential tail. The same resolution functions are used for data and MC. 
Special care is taken to ensure that the simulation takes into account 
the multiple scattering and the time dependence of tracking chamber wire 
inefficiencies and wire resolutions as a function of the drift distance, 
SMD noise and SMD strip inefficiencies. 
Only small final adjustments are needed in order to reflect 
the behaviour of the data. The uncertainty on the adjustment will be
used later for the determination of the systematic error due to resolution
effects. 
Figure~\ref{resf} shows the significance distribution
for all tracks. For negative values of the significance, the good agreement 
between data and MC shows that the resolution 
effects are well understood. The positive part of the significance distribution
is sensitive to the value of $\Rb$. The data shows agreement with the 
MC distribution, which corresponds to a value of $\Rb=0.217$.

\begin{figure}[htbp]
\begin{center}
\includegraphics[width=\textwidth]{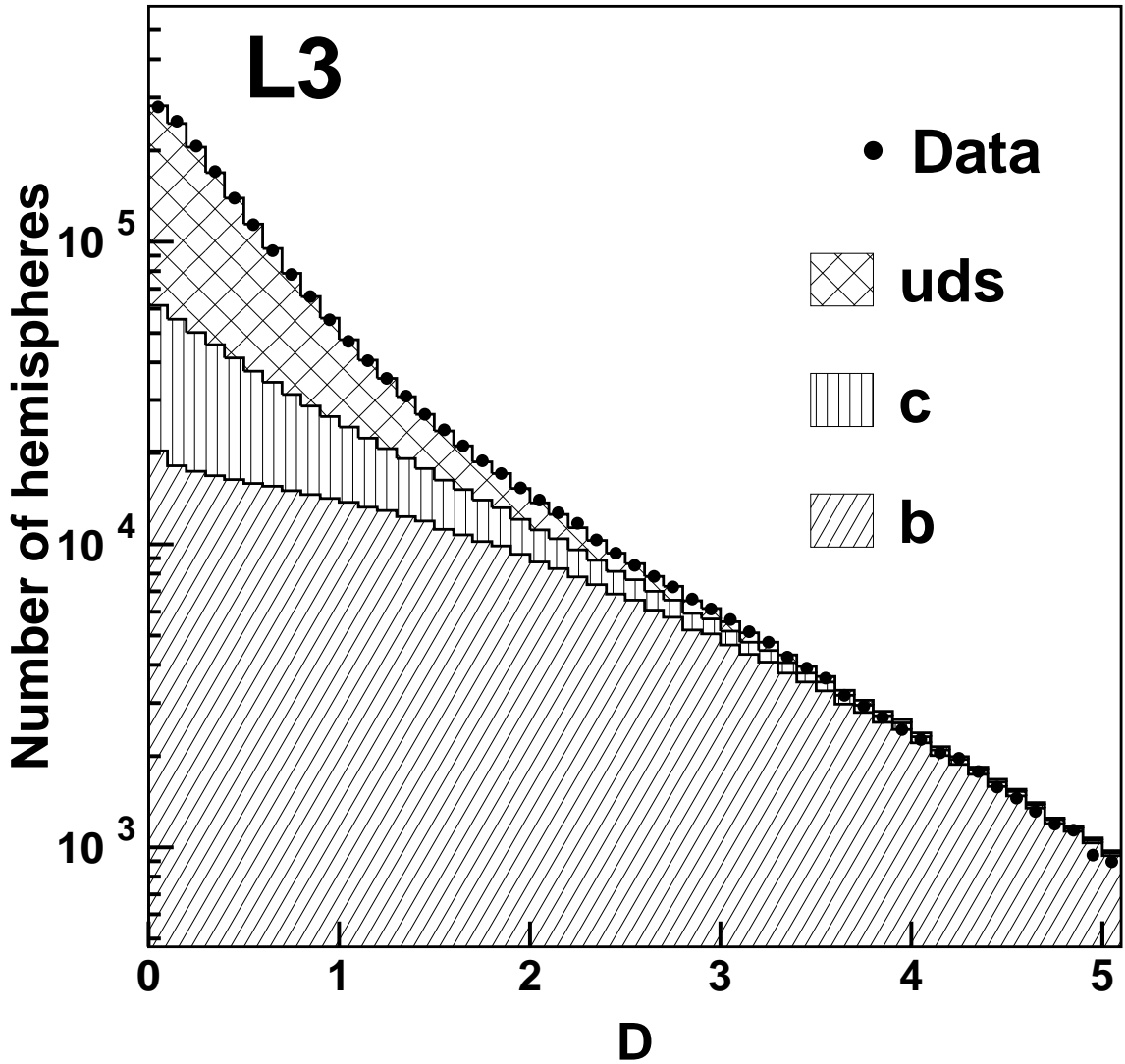}
\end{center}
\caption{\label{discr} Distribution of the discriminant variable, $D$, 
  in data compared
  with the MC prediction. The flavour composition of the 
  hadronic sample is also shown. Good agreement is observed. The 
  higher b-tagging power for large values of the discriminant is 
  exhibited.}
\end{figure}

The resolution function determined from data is used to calculate the
track probability $P(s)$. The distribution
of the hemisphere discriminant $D$ is shown in Figure~\ref{discr}, together
with the MC expectation and its components, in terms of primary quark
flavours. The agreement is satisfactory for the bulk, as well as the tail of the
distribution, and the tagging power of the discriminant is clearly exhibited.

\begin{figure}[htbp]
\begin{center}
\includegraphics[width=\textwidth]{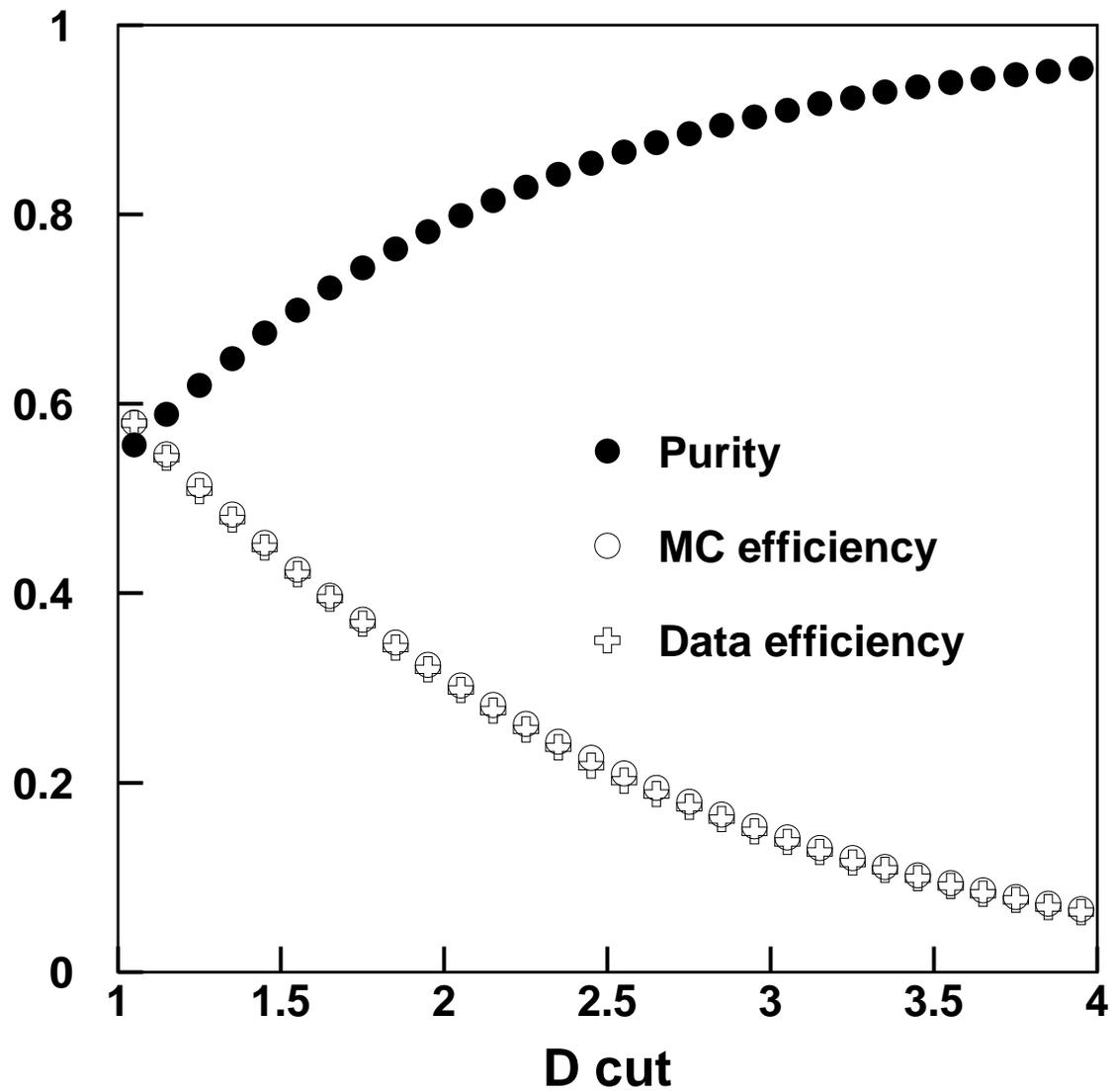}
\end{center}
\caption{\label{effipur} Efficiency and purity of the sample
obtained as a function of the cut on the discriminant variable, $D$. 
For a discriminant cut at $D>2.3$, the purity is 0.843.}
\end{figure}

The efficiency for b-tagging and the purity of the obtained sample can be varied
by changing the cut on the discriminant variable.
We obtain a b-tagging efficiency as a function of
the sample purity as shown in Figure~\ref{effipur} for data and for the MC
simulation.  There is a residual difference between the efficiency
observed in data with respect to the one predicted by MC, which never
exceeds 2.5\%. It is independent of the discriminant cut value over a wide 
range and it is consistent with the estimated statistical and systematic 
errors. For the discriminant cut used in the double-tag analysis ($D>2.3$)
we obtain 
$\epsb^{\rm data} = (23.74 \pm 0.19\stat \pm 0.22\sys)\%$,
whereas the MC estimate is
$\epsb^{\rm MC} = (24.21 \pm 0.03\stat \pm 1.58\sys)\%$, where the 
systematic error is dominated by the b-physics modelling uncertainties.
These uncertainties do not
propagate to an error in $\Rb$ since the MC efficiency is not used in
its determination.

\subsection{Systematic Errors}

\subsubsection{\label{ip:reserr} Tracking Resolution}

The tracking resolution function is determined from data alone. Its statistical 
accuracy is such that it causes a negligible uncertainty on the measurement, 
but a wrong description in the MC simulation influences the values of 
auxiliary parameters like the efficiencies for lighter quarks and the hemisphere 
correlation factors. In order to estimate the systematic error due to uncertainties
in tracking resolution, two MC samples are used, 
one with the final adjustment, ${\rm MC_{final}}$, and the other corresponding to 
the 1-sigma resolution uncertainty, ${\rm MC_{1\sigma}}$. 
The sample ${\rm MC_{final}}$ is found to produce a stable value
of $\Rb$ as a function of the discriminant cut within statistical and 
systematic errors.
${\rm MC_{1\sigma}}$ is defined by the change in the final adjustment that  
leads to a behaviour of $\Rb$ which is inconsistent with a constant value 
by one standard deviation of the observed fluctuations.
The differences in $\epsc$, $\epsuds$ and $\cb$ predicted by the two 
MC samples are propagated as an estimate of the error due to tracking
resolution uncertainties.

\subsubsection{\label{ip:moderr} Systematic Error from Background Modelling}

  MC simulation is needed to determine $\epsuds$, $\epsc$, and $\cb$.
For the charm efficiencies an accurate knowledge of production
and decay properties of the charmed hadrons is important, since the 
different species,
$\Do$, $\Dp$, $\Ds$ and $\Lc$, have
lifetimes varying in the range of 0.2 to 1.1 ps.
Modelling uncertainties in $\epsuds$
arise from the residual contamination by light hadrons with long
lifetime, $\Kos$ and $\Lambda$, as well as the probabilities for
gluon splitting into $\mathrm{b \bar{b}}$ and $\mathrm{c \bar{c}}$ pairs.
Modelling uncertainties of the b-hadron properties only influence the correlation
factor $\cb$. They are estimated by varying the mean value of the B energy fraction 
$<x_E({\rm b})>$, the charged decay multiplicity and the average lifetime of 
b-hadrons.

The variation of the parameters is performed following the suggestions 
of References \cite{lephf,lephf96,lephf98}. The parameter ranges are listed in Table 
\ref{tab:ipsources}. Tables \ref{tab:ipec} and \ref{tab:iped} 
show the complete list of systematic uncertainties on 
$\epsc$ and $\epsuds$ due to the propagation of these modelling 
uncertainties. 
The uncertainties on $\cb$ are discussed in more detail in the 
following Section.

\begin{table}[htbp]
\begin{center}
\begin{tabular}{|l|l|}
  \hline   Error source  &  Variation \\ \hline
\hline
 $\Rc$                                  & $0.1734 \pm 0.0048$ \cite{pdg98} \\
\hline
 Bottom fragmentation parameter:        & \\
  $<x_E({\rm b})>$                      &   $0.709\pm 0.004$ \cite{l3newblife} \\
\hline 
 Bottom decay parameters:               & \\
 B lifetimes                            &  $1.55\pm 0.05$ ps \cite{lephf96} \\
 B decay multiplicity                   &  $4.955 \pm 0.062$ \cite{lephf98} \\
\hline
 Fractions in $\ccbar$ events: & \\
 
 $\Dp$                         &   $0.233\pm 0.027$ \cite{lephf98}  \\
 $\Ds$                         &   $0.103\pm 0.029$ \cite{lephf98}  \\
 $\Lc$                         &   $0.063\pm 0.028$ \cite{lephf98}  \\ \hline
 Gluon splitting in $\ccbar$ events: & \\
 $g \rightarrow \ccbar$          &   $(2.33 \pm 0.50)\%$ \cite{lephf98}  \\
 $g \rightarrow \bbbar$          &   $(0.269\pm 0.067)\%$ \cite{lephf98} \\
\hline 
 Charm decay parameters:                & \\
 $\Do$ lifetime                &   $0.415 \pm 0.004$ ps \cite{pdg98} \\
 $\Dp$  lifetime               &   $1.057 \pm 0.015$ ps \cite{pdg98}\\
 $\Ds$  lifetime               &   $0.467\pm 0.017$ ps \cite{pdg98}\\
 $\Lc$ lifetime                &   $0.206 \pm 0.012$ ps \cite{pdg98}\\ \hline
   D   decay multiplicity:       & \\
 $\Do\rightarrow$ 0 prong      &   $0.054\pm 0.011$ \cite{lephf98} \\
 $\Do\rightarrow$ 4 prong      &   $0.293\pm 0.023$ \cite{lephf98} \\
 $\Do\rightarrow$ 6 prong      &   $0.019\pm 0.009$ \cite{lephf98} \\
 $\Dp\rightarrow$ 1 prong      &   $0.384\pm 0.023$ \cite{lephf98} \\
 $\Dp\rightarrow$ 5 prong      &   $0.075\pm 0.015$ \cite{lephf98} \\
 $\Ds\rightarrow$ 1 prong      &   $0.37\pm 0.10$ \cite{lephf98} \\
 $\Ds\rightarrow$ 5 prong      &   $0.21\pm 0.11$ \cite{lephf98} \\ \hline
 $\mathrm{D} \rightarrow \Kos$ multiplicity &$0.46\pm 0.06$ \cite{markiii,pdg98} \\\hline

 Charm fragmentation parameter:         & \\
  $<x_E({\rm c})>$                      &   $0.484\pm 0.008$ \cite{lephf98} \\
\hline
 Fractions in $\mathrm{uds}$ events:    & \\
 $\Ko$ and $\Lambda$             &   JETSET $\pm$ 10\%        \\
\hline
 Gluon splitting in $\mathrm{uds}$ events: & \\
 $g \rightarrow \ccbar$          &   $(2.33 \pm 0.50)\%$ \cite{lephf98}  \\
 $g \rightarrow \bbbar$          &   $(0.267\pm 0.067)\%$ \cite{lephf98} \\\hline

\end{tabular}
\end{center}
\caption{\label{tab:ipsources} 
Variation of modelling  parameters used for the determination of  the
systematic error in the impact parameter double-tag measurement.}
\end{table}

\subsubsection{\label{ip:corerr} Systematics from Hemisphere Correlations}

   Systematic errors on $\cb$ are due to uncertainties in the MC
simulation. In addition to resolution and modelling effects, taken into 
account as described in Sections \ref{ip:reserr} and \ref{ip:moderr}, reconstruction 
algorithms and detector inhomogeneities may create correlations
between the tagging efficiencies of both hemispheres. 

A possible source of correlation can be quantified by choosing a
variable $\lambda$ for each hemisphere which could be influenced by tagging
the opposite hemisphere. For a particular cut on the hemisphere
discriminant we then define three distributions:

\begin{itemize}
\item The normalised distribution of $\lambda$ for all hemispheres, $N(\lambda)$.
\item The single-hemisphere tagging efficiency as a function of $\lambda$,
  $\epsilon(\lambda)$.
\item The normalised distribution of $\lambda$ in a co-tagged hemisphere,
  $C(\lambda)$. A co-tagged hemisphere is the one opposite to a tagged hemisphere,
  regardless of whether it is itself tagged.
\end{itemize}

We then form a coefficient, $\cb^{\lambda}$, reflecting the correlation
characterised by $\lambda$ for a particular discriminant cut:

\begin{equation}
  \cb^{\lambda} = \frac{\int{\epsilon(\lambda)~C(\lambda)~d\lambda}}
                    {\int{\epsilon(\lambda)~N(\lambda)~d\lambda}}.
\end{equation}

The value $\cb^{\lambda}=1$ implies that there is no correlation between
hemispheres from effects characterised by $\lambda$. A value $\cb^{\lambda}>1$
indicates a positive correlation, $\cb^{\lambda}<1$ an anti-correlation. 

The analysis is performed on several candidate variables $\lambda$.  For a
$\bbbar$ MC sample, the sum of the separate components
is then compared to the total correlation factor
$\cb$. A reasonable agreement between them means that the relevant sources 
of correlation have been identified. The following sources are found to 
be relevant:

\begin{itemize}
\item { Angular effects:} Inefficient regions of the detector can lead
  to correlations due to the back-to-back nature of hadronic events.
  This is estimated using $\lambda=\cos{\theta}$ and $\lambda=\phi$, where $\theta$
  and $\phi$ are the polar and azimuthal angles of the most energetic
  jet in each hemisphere.

\item { Vertex effects:} Both hemispheres use the same primary
  vertex, which is determined as explained in Section \ref{TraVert}.
  The main effect arises because the primary vertex fit may also 
  include tracks coming from b-hadron decays (negative correlation). 
  To quantify the effect, two independent primary vertices are
  constructed separately using the tracks assigned to each hemisphere.
  The variable $\lambda$ is taken to be the distance in the $x-y$ 
  plane between the vertex in each hemisphere and the overall event vertex. Its sign
  is given according to how far each hemisphere vertex
  moves when the beam-spot position constraint is removed: a 
  positive sign is assigned to the hemisphere with the larger movement.

\item { QCD effects:} The presence of hard gluons in the event can
  influence the tagging efficiency of both hemispheres by taking
  energy away from the primary quarks (positive correlation) or, in 
  an extreme case, by pushing both quarks into the same hemisphere
  (negative correlation). This effect is modified by reconstruction 
  and detector resolution effects.
  The signed event thrust, $\lambda = \pm T$, is used as a probe.
  A positive sign is assigned to the hemisphere with the higher energy jet.
\end{itemize}

The dominant components are those due to gluon
radiation and vertex bias. There is a remaining discrepancy between the linear
sum of the correlation components due to the above three sources and the total 
observed correlation factor. That is expected, due to the interference between 
the sources 
considered and to additional sources of less relevance. For instance, $\phi$ 
and vertex effects are intrinsically correlated, since the primary vertex 
uncertainty is affected by the $\phi$-dependent beam spot size.

 To test the quality of the MC correlation simulation,
a 70\% b-purity sample is selected in data by requiring the
event discriminant to be greater than 1.5. The same cut is applied
to the MC sample. The correlation coefficients are calculated for each of the three 
sources in data and MC. The differences are taken as the systematic 
uncertainties on the correlation term and propagated 
through to a systematic error on $\Rb$. The complete list of systematic 
uncertainties on $\cb$ is shown in Table \ref{tab:ipcb}.

\begin{table}[htpb]
\begin{center}
%
%
\begin{tabular}{|l|r|}
  \hline   Error source                           & $\Delta \epsc $   \\
\hline
\hline
 MC Statistics                                    & $0.02\%$               \\
 Track Resolution                                 & $0.02\%$               \\
\hline
 $\Dp$ fraction                          & $+0.10\%$               \\
 $\Ds$ fraction                          & $+0.01\%$               \\
 $\Lc$ fraction                          & $-0.03\%$               \\
\hline
 $\Do$ lifetime                          & $+0.02\%$               \\
 $\Dp$ lifetime                          & $+0.01\%$               \\
 $\Ds$ lifetime                          & $+0.01\%$               \\
 $\Lc$ lifetime                          & $+0.01\%$               \\
\hline
 Decay multiplicities:                & \\
 $\Dp$ 1-prong         & $-0.03\%$               \\
 $\Dp$ 5-prong         & $ 0.00\%$               \\
 $\Do$ 0-prong         & $ 0.00\%$               \\
 $\Do$ 4-prong         & $+0.02\%$               \\
 $\Do$ 6-prong         & $ 0.00\%$               \\
 $\Ds$ 1-prong         & $-0.02\%$               \\
 $\Ds$ 5-prong         & $-0.04\%$               \\
\hline
 $\mathrm{D} \rightarrow \Kos$ multiplicity      & $+0.02\%$               \\
 \hline
 $<x_E({\rm c})>$                                 & $+0.06\%$               \\
 \hline
 $g \rightarrow \ccbar$                    & $0.00\%$               \\       
 $g \rightarrow \bbbar$                    & $0.00\%$               \\       
 \hline
 \hline
 Total                                            & $0.14\%$               \\
\hline

\end{tabular}
\end{center}
\caption{\label{tab:ipec} 
    Error contributions to $\epsc$ for a cut at $D > 2.3$. 
    The sign associated to a given error indicates 
    the effect of a positive variation of the corresponding parameter in 
    Table \ref{tab:ipsources}.}

\end{table}
\begin{table}[htpb]

\begin{center}
\vspace{.5 cm}
\begin{tabular}{|l|r|}
  \hline   Error source  &$\Delta \epsuds $ \\ 
\hline
\hline
 MC Statistics           &  $0.00\%$                 \\ 
 Track Resolution        &  $0.01\%$                 \\ 
\hline
 K$^0$ and Hyperons      & $+0.03\%$                 \\ 
\hline 
$g \rightarrow \ccbar$ & $0.00\%$                 \\  
$g \rightarrow \bbbar$ & $0.00\%$                 \\  
 \hline
 \hline
Total                    &  $0.03\%$                 \\ \hline
\end{tabular}
\end{center}
\caption{\label{tab:iped} Error contributions to $\epsuds$
for a cut at $D > 2.3$.
    The sign associated to a given error indicates 
    the effect of a positive variation of the corresponding parameter in 
    Table \ref{tab:ipsources}.}
\end{table}
%
%

\begin{table}[htpb]

\begin{center}
\vspace{.5 cm}
\begin{tabular}{|l|r|}
\hline   Error source  &$\Delta \cb$ \\ 
\hline
\hline
 MC Statistics           &  $0.0036$                \\
 Track Resolution        &  $0.0002$                \\
 Vertex bias             &  $0.0053$                \\
 $\theta$ correlations   &  $0.0002$                 \\
 $\phi$ correlations     &  $0.0009$                 \\
 Hard gluon emission     &  $0.0004$                \\
\hline
 B fragmentation         & $+0.0014$                \\
\hline
 B lifetimes             & $+0.0008$                \\ 
 B decay multiplicity    & $+0.0004$                \\ 
 \hline
 \hline
Total                    &  $0.0067$                \\ \hline
\end{tabular}
\end{center}
\caption{\label{tab:ipcb}
   Error contributions to $\cb$.
    The sign associated to a given error indicates 
    the effect of a positive variation of the corresponding parameter in 
    Table \ref{tab:ipsources}.}
\end{table}

\subsection{Results}
\begin{figure}[htbp]
\begin{center}
\includegraphics[width=\textwidth]{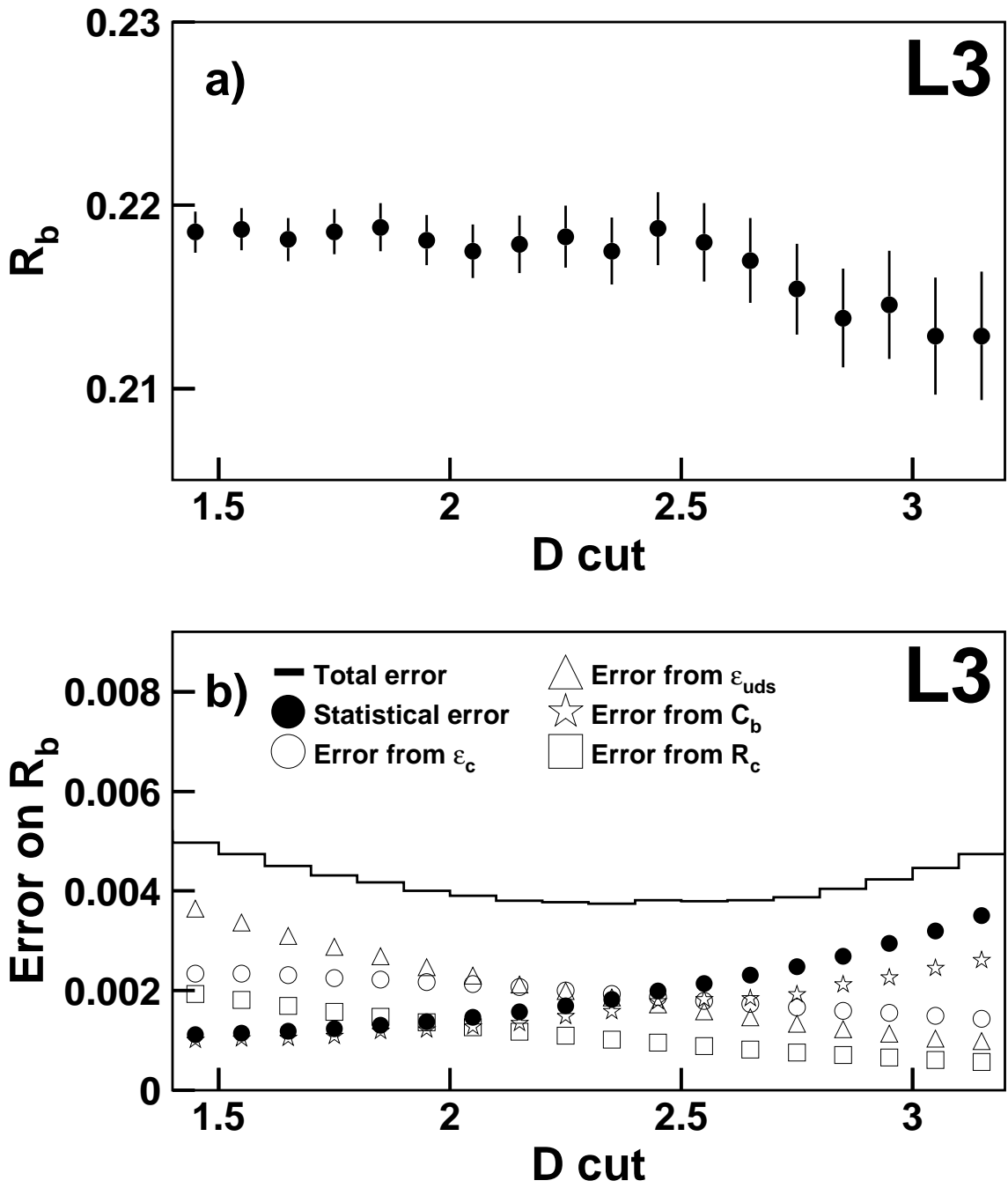}
\end{center}
\caption{\label{rb} a) Value of $\Rb$ obtained as a function of the 
cut on the impact parameter discriminant, $D$. b) Statistical, 
systematic and total errors 
for $\Rb$ as a function of the discriminant cut. The uncertainty on $\Rc$, 
$\Delta \Rc = 0.0048$, is taken from Reference \protect\cite{pdg98}.}
\end{figure}

  Systematic errors on $\epsc$, $\epsuds$ and
$\cb$, are propagated into the double-tag 
$\Rb$ measurement. Figure~\ref{rb}a) shows the measured value of $\Rb$ 
as a function of the discriminant cut in the region around the minimum 
of the total uncertainty. The value is stable within the estimated statistical and 
systematic errors. Figure~\ref{rb}b) shows the
statistical and systematic errors 
on $\Rb$ in the same range. The minimum of the total error occurs at
$D>2.3$, which defines the 
central value of our measurement. For this cut, 118817 tagged hemispheres and
11705 double-tagged events are selected from a sample 
of 968964 hadronic events.

 The value of $\epsc$ is estimated by MC to be
$\epsc = (3.05 \pm 0.02\stat \pm 0.14\sys)\%$.
A breakdown of the error is shown in Table \ref{tab:ipec}.
Since  the individual charm lifetimes are measured  very accurately,
the fractions of the different species are the major error contributions. 
Among all the charmed hadrons the $\Dp$
properties lead to the dominant errors because it has the longest lifetime.

The value of $\epsuds$ is estimated to be
$\epsuds = (0.739 \pm 0.004\stat \pm 0.035\sys)\%$.
The different error contributions are shown in Table \ref{tab:iped}.
The systematic error is dominated by the uncertainty in the rate of 
light-flavoured hadrons with long lifetime.

The value of the correlation coefficient is
$\cb = 0.9717 \pm 0.0036\stat \pm 0.0056\sys$. The different error
contributions are listed in Table \ref{tab:ipcb}. Primary vertex effects are 
the dominant source of uncertainty.

The measured values of $\Rb$ and $\epsb$ are:
\begin{eqnarray}
 \Rb & = &  0.2173 \pm 0.0018\stat \pm 0.0032\sys, \\
 \epsb & = & (23.74 \pm 0.19\stat \pm 0.22\sys)\%.
\end{eqnarray}

The detailed list of contributions to the error on $\Rb$ is given in 
Table \ref{tab:iprb}.
The sources internal to L3 are separated from the ones
in common with other experiments. 

\begin{table}[htpb]
\begin{center}
%
%
\begin{tabular}{|l|r|}
\hline   
 \multicolumn{2}{|c|}{$\Delta \Rb$ from Internal Error Sources} \\
\hline
\hline
 MC statistics                               & $0.00092$ \\
 Resolution                                  & $0.00056$ \\
 Vertex effects on $\cb$                     & $0.00125$ \\
 $\theta$ effects on $\cb$                   & $0.00006$ \\
 $\phi$ effects on $\cb$                     & $0.00021$ \\
\hline
\hline
 Total Internal                              & $0.00166$ \\
\hline 
\hline   
 \multicolumn{2}{|c|}{$\Delta \Rb$ from External Error Sources} \\
\hline   
\hline   
 $\Rc$ uncertainty                  &   $-0.00094$ \\
\hline   
 $\Dp$ fraction                     &   $-0.00128$ \\
 $\Ds$ fraction                     &   $-0.00017$ \\
 $\Lc$ fraction                     &   $+0.00045$ \\
\hline
 $\Dp$ lifetime                     &   $-0.00020$ \\
 $\Do$ lifetime                     &   $-0.00013$ \\
 $\Ds$ lifetime                     &   $-0.00012$ \\
 $\Lc$ lifetime                     &   $-0.00008$ \\
\hline
 $\Dp$ 1-prong decay multiplicity   &   $+0.00045$ \\
 $\Dp$ 5-prong decay multiplicity   &   $-0.00006$ \\
 $\Do$ 0-prong decay multiplicity   &   $-0.00004$ \\
 $\Do$ 4-prong decay multiplicity   &   $-0.00024$ \\
 $\Do$ 6-prong decay multiplicity   &   $ 0.00000$ \\
 $\Ds$ 1-prong decay multiplicity   &   $+0.00028$ \\
 $\Ds$ 5-prong decay multiplicity   &   $+0.00055$ \\
\hline
 $\mathrm{D} \rightarrow \Kos$ multiplicity  &   $-0.00025$ \\
\hline
 $<x_E({\rm c})>$                            &   $-0.00086$ \\
\hline
 $g \rightarrow \ccbar$ in $\ccbar$ events  
                                             &   $-0.00001$ \\
 $g \rightarrow \bbbar$ in $\ccbar$ events  
                                             &   $-0.00003$ \\
\hline
 $K^0_s$ and $\Lambda$ production            &   $-0.00176$ \\
\hline
 $g \rightarrow \ccbar$ in $\mathrm{uds}$ events  
                                             &   $-0.00018$ \\       
 $g \rightarrow \bbbar$ in $\mathrm{uds}$ events
                                             &   $-0.00015$ \\
\hline
 B fragmentation $<x_E({\rm b})>$            &  $+0.00032$ \\
\hline
 B lifetimes                                 &  $+0.00018$ \\
 B decay multiplicity                        &  $+0.00009$ \\
 Hard gluon fragmentation                    &  $ 0.00010$ \\
\hline 
\hline
 Total External                              &   $0.00271$ \\
\hline

\end{tabular}
\end{center}
\caption{\label{tab:iprb} 
   Internal and external error contributions to $\Rb$ for a cut at $D > 2.3$.
   The errors from the D mesons fractions are propagated according to the 
   correlation matrix defined in Reference \protect\cite{lephf98}.
   The sign associated to a given error indicates 
   the effect of a positive variation of the corresponding parameter in 
   Table \ref{tab:ipsources}.}
\end{table}

\section{Lepton Analysis} \label{sec:leptons}

   The selection of leptons with high momentum and high transverse momentum 
with respect to the closest jet is applied to the set of events passing the 
hadronic selection. It requires a good performance of the central tracking 
detectors and a restriction to the angular region: $|\cos{\theta_T}| < 0.7$.

\subsection{Lepton Identification}

An electron is identified as a cluster in the electromagnetic calorimeter 
matched to a track in the central tracking chamber. The requirements are:

\begin{itemize}

\item The cluster in the electromagnetic calorimeter must have more than 
      5 crystals with signal, more than 90\% of the energy contained
      in a $3\times 3$ matrix and a shower shape consistent with the 
      deposit of an electromagnetic particle.

\item The cluster energy must be greater than $3 \GeV$.

\item The cluster  must be matched in azimuthal angle (5 mrad) and 
      in energy (within 4 standard deviations) to a track in the central 
      tracking chamber. 

\item The energy deposited in the hadron calorimeter in a cone of half 
      opening angle of $7^{\circ}$ around the cluster direction must be less 
      than $3 \GeV$.

\end{itemize}

\noindent A muon is identified as a track in the muon chambers which satisfies
the following requirements:

\begin{itemize}

\item The muon track is required to have track segments in at least two of the three
      layers of the chambers which measure the trajectory in the plane transverse 
      to the beam. In addition, it must have at least one of the two hit segments
      in the muon chambers which measure longitudinal coordinates.

\item The measured muon momentum must be greater than $3 \GeV$.

\item The track must point back to the interaction vertex region within 3 standard
      deviations of the uncertainty in the extrapolation. This uncertainty 
      takes into account multiple scattering and energy losses in the calorimeters.

\item The muon track must be matched to a track in the central tracking chamber.

\end{itemize}

\subsection{\boldmath Heavy-Quark Tagging Using High-$\pt$ Leptons}

\begin{figure}[htbp]
\begin{center}
\includegraphics[width=\textwidth]{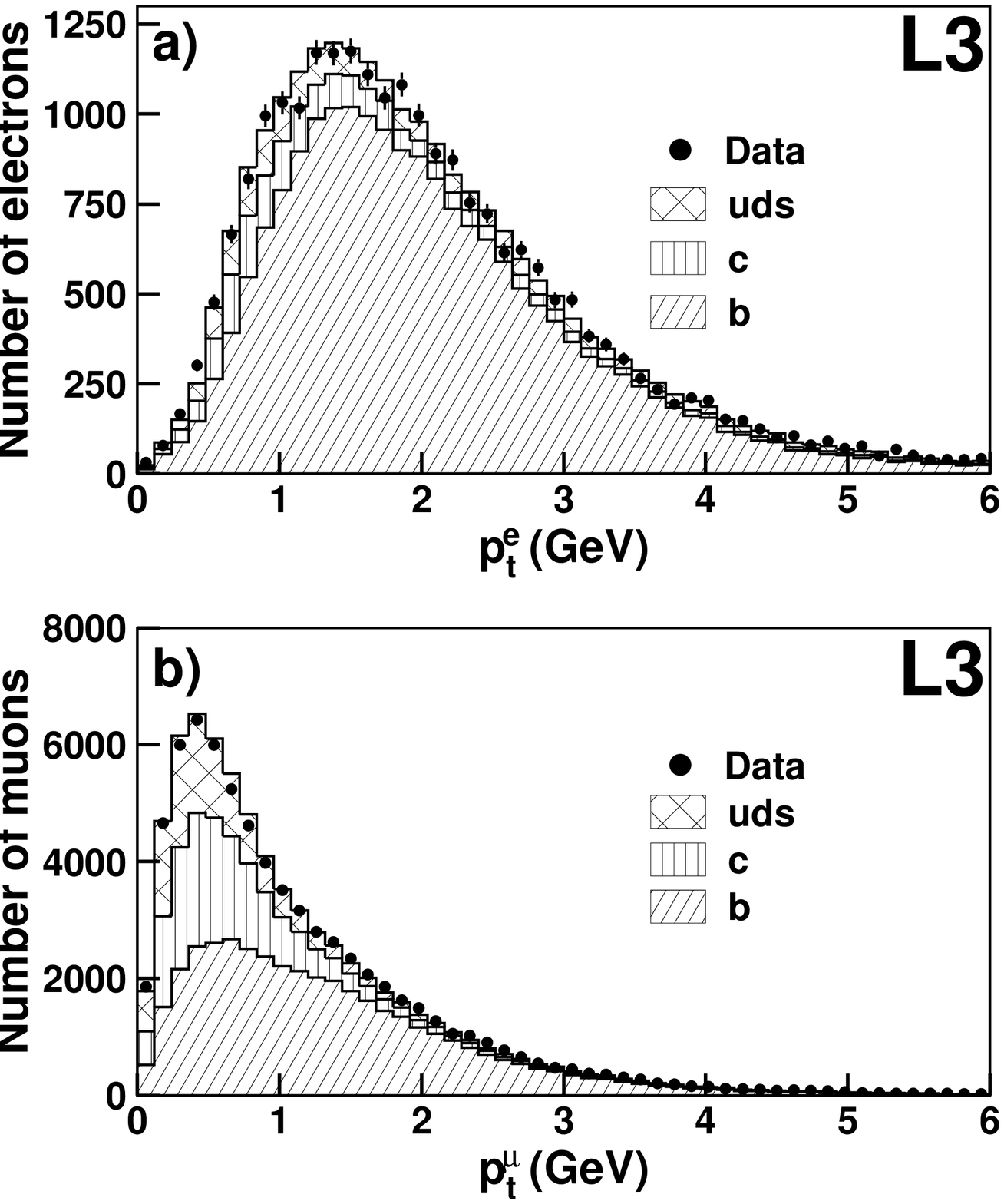}
\end{center}
\caption{\label{fig:ptlept} 
Distributions of the transverse momentum with respect to the closest 
jet for electrons a) and muons b). The transverse momentum of the lepton 
$\ell$ is denoted by $\pt^{\ell}$. The histogram 
shows the flavour composition of the MC sample.}
\end{figure}

   Hemispheres containing high-energy leptons are selected
as b-candidates. Due to the hard b-fragmentation and the large b-mass, a higher 
b-purity sample
is obtained as the transverse momentum of the lepton with respect to the 
b-jet, $\pt$, increases. The distribution of this variable is shown in Figure 
\ref{fig:ptlept}. Very good agreement between data and simulation is observed. 
A $\pt$ cut of $1 \GeV$ is used for the double-tag measurement.

\subsection{Systematic Errors}

\subsubsection{Lepton Selection Uncertainties}

   Uncertainties in the lepton selection are expected if 
electromagnetic clusters, punch-through and tracking resolutions are not well 
reproduced in the simulation. Since the MC simulation takes into 
account the behaviour 
of the detector as a function of time, only a small uncertainty is expected. 
It is estimated by studying the statistical consistency of the 
measured $\Rb$ values as a function of the $\pt$ cut.

\subsubsection{\label{le:moderr}Systematics from Background Modelling}

   The charm efficiencies depend on the assumptions for the overall rate 
and properties of semileptonic c-decays. The uncertainty on these 
assumptions is taken into account by varying the semileptonic branching 
fraction, $\Brcl$, the charm fragmentation, $<x_E({\rm c})>$, and the 
semileptonic decay model within the ranges suggested in Reference 
\cite{lephf98}. 
The parameter ranges are summarised in Table \ref{tab:lesources}.
   The uncertainty on $\epsuds^\prime$ is estimated by 
selecting a sample enriched in light quarks.
Both hemispheres in the event must satisfy a cut on the discriminant variable
$D<0.5$. In this sample a maximum discrepancy of $3\%$ is found between 
data and MC for the number of leptons with momentum below $5 \GeV$. 
This discrepancy is translated into a relative error in $\epsuds^\prime$.

\begin{table}[htbp]
\begin{center}
\begin{tabular}{|l|l|}
  \hline   Error source  &  Variation \\ 
\hline \hline
 $\Rc$ uncertainty                      & $0.1734 \pm 0.0048$ \cite{pdg98} \\
\hline
 Charm fragmentation parameter:         & \\
  $<x_E({\rm c})>$                      &   $0.484\pm 0.008$ \cite{lephf98} \\
\hline 
 $\Brcl$                                &   $(9.8 \pm 0.5)\%$ \cite{lephf98} \\
\hline 
 Semileptonic Decay Model ${\rm c} \rightarrow \ell$ & 
           \begin{tabular}{c}
                   ${\rm +ACCMM~model~1}$ \\ ${\rm -ACCMM~model~2}$ 
           \end{tabular}
                   \cite{lephf} \\
\hline
 Light quark uncertainties              &  Data-MC comparisons \\
\hline
\end{tabular}
\end{center}
\caption{\label{tab:lesources} Variation of modelling  parameters used for the 
determination of  the systematic error in the lepton double-tag measurement.}
\end{table}

\subsubsection{Systematics from Hemisphere Correlations}

    The correlation between the semileptonic decay properties of the two 
hemispheres is expected to be very small. The value obtained for $\cb^\prime$ 
from the MC simulation is $\cb^\prime = 0.993 \pm 0.009$, consistent with 
no correlation.

    Evidence of non-negligible correlation effects have been looked for. The 
following quantity is defined:

\begin{equation}
 C = \frac{<x_1 x_2>}{<x_1>~<x_2>},
\end{equation}

\noindent
where $x_1$ and $x_2$ are values of physical variables in the two hemispheres 
of the event. We have used the lepton momentum and transverse momentum, since 
they are the relevant variables of the analysis. The values obtained for $C$ 
are consistent with unity within the statistical errors, independent of
the $\pt$ cut applied and of restrictions in the detector acceptance.
 Since no evidence of correlation is found, only the MC statistical
uncertainty is assigned as a systematic error.

\subsection {Results}

   For a $\pt$ cut of $1 \GeV$, 49308 hemispheres are tagged and
1927 events are double-tagged on a sample of 968964 hadronic events.

  The value of $\epsc^\prime$ 
is found to be $\epsc^\prime = (1.70 \pm 0.01\stat \pm 0.11\sys)\%$.
The error is dominated by the background modelling of semileptonic 
charm decays. The value of $\epsuds^\prime$ is determined to be
$\epsuds^\prime = (0.362 \pm 0.003\stat \pm 0.011\sys)\%$,
and the value of the correlation coefficient is 
$\cb^\prime = 0.993 \pm 0.009\stat$. 

  Taking into account statistical and systematic errors we obtain:
\begin{eqnarray}
\Rb & = & 0.2116 \pm 0.0050\stat \pm 0.0045\sys, \\
\epsb^\prime & = & (9.59 \pm 0.22\stat \pm 0.11\sys) \%.
\end{eqnarray}

  The different contributions to the systematic error on $\Rb$ are shown in Table 
\ref{tab:lerb}.

\begin{table}[htbp]
\begin{center}
\vspace{.5 cm}
\begin{tabular}{|l|r|}
\hline
  Sources of systematics & Error on $\Rb$ \\ 
\hline
\hline
   MC statistics              & $ 0.0022$ \\
   Lepton Identification      & $ 0.0005$ \\
\hline
   $\Rc$                      & $-0.0013$ \\
\hline
   $\Brcl$                    & $-0.0028$ \\
\hline
   Semileptonic Decay Model   & $-0.0022$ \\
\hline
   $<x_E({\rm c})>$           & $-0.0006$ \\
\hline
   Light quark uncertainties  & $-0.0007$ \\
\hline
\hline
  Total  & 0.0045 \\ 
\hline
\end{tabular}
\end{center}
\caption{\label{tab:lerb} Systematic error contributions to $\Rb$ from
          the lepton double-tag measurement.
          The sign associated to a given
          error indicates the effect of a positive variation of the 
          corresponding parameter in Tables \ref{tab:ipsources} and 
          \ref{tab:lesources}.}
\end{table}

\begin{table}[htpb]
\begin{center}
%
%
\begin{tabular}{|l|r|}
\hline   
 \multicolumn{2}{|c|}{$\Delta \Rb$ from Internal Error Sources} \\
\hline
\hline
 MC statistics                               & 0.00081  \\ 
 Tracking Resolution                         & 0.00042  \\
 Vertex effects on $\cb$                     & 0.00079  \\
 $\theta$ effects on $\cb$                   & 0.00003  \\
 $\phi$ effects on $\cb$                     & 0.00013  \\
 Lepton identification                       & 0.00010  \\
 Correlation between tags                    & 0.00079  \\
\hline
 Total Internal                              & 0.00145  \\
\hline
\hline
 \multicolumn{2}{|c|}{$\Delta \Rb$ from External Error Sources} \\
\hline
\hline
 $\Rc$                              &   $-0.00104$ \\
\hline
 $\Dp$ fraction                     &   $-0.00100$ \\
 $\Ds$ fraction                     &   $-0.00013$ \\
 $\Lc$ fraction                     &   $+0.00035$ \\
\hline
 $\Dp$ lifetime                     &   $-0.00016$ \\
 $\Do$ lifetime                     &   $-0.00010$ \\
 $\Ds$ lifetime                     &   $-0.00010$ \\
 $\Lc$ lifetime                     &   $-0.00006$ \\
\hline
 $\Dp$ 1-prong decay multiplicity   &   $+0.00035$ \\
 $\Dp$ 5-prong decay multiplicity   &   $-0.00005$ \\
 $\Do$ 0-prong decay multiplicity   &   $-0.00003$ \\
 $\Do$ 4-prong decay multiplicity   &   $-0.00018$ \\
 $\Do$ 6-prong decay multiplicity   &   $ 0.00000$ \\
 $\Ds$ 1-prong decay multiplicity   &   $+0.00022$ \\
 $\Ds$ 5-prong decay multiplicity   &   $+0.00043$ \\
\hline
 $\mathrm{D} \rightarrow \Kos$ multiplicity  &   $-0.00019$ \\
\hline
 $<x_E({\rm c})>$                            &   $-0.00082$ \\ 
 $\Brcl$                                     &   $-0.00068$ \\
 Semileptonic Decay Model                    &   $-0.00054$ \\
\hline
 $g \rightarrow \ccbar$ in $\ccbar$ events  
                                             &   $-0.00001$ \\
 $g \rightarrow \bbbar$ in $\ccbar$ events  
                                             &   $-0.00002$ \\
\hline
 $K^0_s$ and $\Lambda$ in uds events         &   $-0.00131$ \\
 Light quark uncertainties                   &   $-0.00007$ \\
\hline
 $g \rightarrow \ccbar$ in $\mathrm{uds}$ events  
                                             &   $-0.00013$ \\       
 $g \rightarrow \bbbar$ in $\mathrm{uds}$ events  
                                             &   $-0.00011$ \\
\hline
 B fragmentation $<x_E({\rm b})>$            &  $+0.00020$ \\
 B lifetimes                                 &  $+0.00011$ \\
 B decay multiplicity                        &  $+0.00006$ \\
 Hard gluon fragmentation                    &  $ 0.00007$ \\
\hline
\hline
 Total External                              &   0.00244 \\
\hline

\end{tabular}
\end{center}
\caption{\label{tab:allsys} 
   Internal and external error contributions to the $\Rb$ measurement.
   The errors from the D meson fractions are propagated according to the 
   correlation matrix defined in Reference \protect\cite{lephf98}. 
   Other errors in the table can be considered as uncorrelated.
          The sign associated to a given
          error indicates the effect of a positive variation of the 
          corresponding parameter in Tables \ref{tab:ipsources} and 
          \ref{tab:lesources}.}
\end{table}

\section{\boldmath Measurement of $\Rb$}
  For an impact parameter discriminant cut of 
$D>2.3$ and a lepton $\pt$ cut of $1 \GeV$, 9368 events are selected with a 
lepton tag and an impact parameter tag in opposite hemispheres.
The correlation factor between the tags is 
$\cb^{\prime\prime} = 1.004 \pm 0.005\stat \pm 0.015\sys$, 
where the systematic error takes into account detector and
reconstruction effects. 

    A combined fit using Equations \ref{eq1}-\ref{eq5} for the
impact parameter and lepton tags is performed. The MC input efficiencies
and correlation factors with their uncertainties are included as constraints 
in the fit. The final results are:
\begin{eqnarray}
\Rb & = & 0.2174 \pm 0.0015\stat \pm 0.0028\sys, \\
\epsb & = & (23.70 \pm 0.15\stat \pm 0.19\sys)\%, \\
\epsb^\prime & = & (9.34 \pm 0.07\stat \pm 0.10\sys)\% \label{eprimeb}.
\end{eqnarray}

 The statistical error of the measurement is determined by fixing the MC
efficiencies and correlations to their fitted values. The systematic 
uncertainties are propagated according to the covariance matrix of the fit.
Charm, light-quark and b-quark systematics are uncorrelated, except for the 
case of the charm fragmentation, which correlates the impact parameter and lepton 
tags.  All systematic error contributions to the measurement are shown in Table 
\ref{tab:allsys}. 

  The measurement of $\Rb$ is in agreement with the expectation
from the Standard Model, $\Rb^{SM} = 0.2158 \pm 0.0002$ \cite{zfitter},
and is compatible with previous determinations \cite{otherlep}.

\section{\boldmath Measurement of {\boldmath $\Brbl$}}

  The efficiency $\epsb^\prime$, given by Equation \ref{eprimeb},
 quantifies the fraction of high $p,\pt$ leptons in b-jets. Therefore, it is 
sensitive to the value of the semileptonic branching ratio of b-hadrons at LEP.
In the reference MC with $\Brbl = 10.45\%$ we find:
\begin{eqnarray}
 \epsb^{\prime~{\rm REF}} = (9.50 \pm 0.04\stat \pm 0.11\sys)\%.
                                                      \label{eprimebref}
\end{eqnarray}

The systematic error on $\epsb^\prime$ is almost independent 
of b-quark model assumptions, since double-tag methods are used. 
The dependence on $\Rb$ is experimental, due to the simultaneous determination 
of both parameters (the statistical correlation is $-0.72$).
 The central value and the systematic uncertainty 
are determined for the set of parameters and variations shown in Table
\ref{tab:epsilonbl_mc}. 
The dominant systematic errors in $\epsb^{\prime~{\rm REF}}$ are those 
due to detector inefficiencies and to 
semileptonic decay modelling uncertainties. They will be discussed in the next 
subsections. 

  For different values of the branching ratio we find a linear dependence of 
$\epsb^\prime$: 
\begin{equation}
\epsb^{\prime} = \epsb^{\prime~{\rm REF}} + 0.5444~[\Brbl-\Brbl^{\rm REF}]. \label{eq:brl_formula}
\end{equation}

\subsection{Detector Inefficiencies}
Time-dependent inefficiencies of the L3 detector are expected to be reproduced 
in the simulation within a few percent accuracy. In order to estimate possible 
extra sources of uncertainty not taken into account 
in the MC, we have used large data samples of $\ee \ra \ee$ and $\ee \ra \mm$.
The samples are selected by requiring at least one lepton with high energy 
($> 0.35~\rts$) and applying the same cuts used for the 
sample described in Section \ref{sec:leptons}, except for the transverse 
momentum cut. The ratio of the number of one-lepton
events to the number of two-lepton events is determined for both 
data and MC including backgrounds. The comparison suggests extra
inefficiencies of $(1.8 \pm 0.1)\%$ and $(3.2 \pm 0.1)\%$ for the electron
and muon samples, respectively. The errors are only statistical. These numbers
are determined under the assumption that inefficiencies in both hemispheres of 
the event are uncorrelated. This is confirmed by the agreement at the percent 
level between the total cross sections measured in data and MC, which can be
largely affected by simultaneous losses of the two leptons in an event.

   Additional checks at lower lepton energy 
($\tau \ra \ell \nu_\tau \overline{\nu}_\ell$, $\ee \ra \ee \ll$) confirm the
extra inefficiencies observed at high energy.
Since the discrepancy is larger for muons, an additional cross check 
is performed. A sample is selected on the basis of the expected deposition in 
the hadron calorimeter for minimum ionising particles. The percentage of muons 
found in this sample is compared to MC, leading
to an estimated extra inefficiency of $(3.5 \pm 0.5)\%$. When 
restricted to muons of low energy, we obtain $(3.2 \pm 1.2)\%$.

As a consequence of this study the following relative losses 
have been added in the simulation: $(1.8 \pm 1.0)\%$ 
for electrons and $(3.2 \pm 1.0)\%$ for muons. The reference value in Equation 
\ref{eprimebref} already 
takes into account these acceptance corrections.

\begin{table}[htpb]
\begin{center}
\begin{tabular}{|l|c|r|}
\hline   
 Parameter & Variation & Change in $\epsb^{\prime}$ (\%) \\
\hline
\hline
 $\mathrm{e}$ Detector efficiency & $\pm 1\%$ & $+0.037$ \\
 $\mu$ Detector efficiency & $\pm 1\%$ & $+0.058$ \\
 Decay parameters in $b \ra \ell$ & $\pf= 286 \pm 35 \MeV$ 
                           & $-0.080$ \\
 Decay scheme in $b \ra \ell$ & $\pm ({\rm ISGW}^{**} - {\rm ACCMM})$ 
                           & $-0.060$ \\
 $<x_E({\rm b})>$ & $0.709 \pm 0.004$ \cite{l3newblife} & $+0.032$ \\
 $\Brbcl$ & $(8.09 \pm 0.5)\%$ \cite{lephf96} & $-0.005$ \\
 $\Brbcbarl$ & $(1.66 \pm 0.40)\%$ \cite{lephf98} & $-0.010$ \\
 b-lifetime & $1.55 \pm 0.05$ ps \cite{lephf96} & $0.000$ \\
 $\Brbtaul$ & $(0.452 \pm 0.074)\%$ \cite{lephf98} & $-0.003$ \\
 $\Brbjl$ & $(0.07 \pm 0.02)\%$ \cite{lephf98} & $-0.006$ \\
\hline
\hline
 Total & & 0.11  \\
\hline
\end{tabular}
\end{center}
\caption{\label{tab:epsilonbl_mc} 
  Central values assumed for the MC determination of 
$\epsb^\prime$.
  Systematic errors and ranges of variation are also shown.
          The sign associated to a given
          error indicates the effect of a positive variation of the 
          parameter.}
\end{table}

\subsection{Modelling Uncertainties}

   The largest source of systematic error comes from the uncertainties in 
the modelling of the $\blnuX$ decay. The MC simulation uses a model
which can be approximately described by the ACCMM \cite{accmm} 
parametrisation with a c-quark
mass $\mc = 1.67 \GeV$, a Fermi momentum inside the B meson
$\pf = 298 \MeV$, and a mass of the spectator quark $\msp = 150 \MeV$. 
In order to study the sensitivity to modelling assumptions, a sample enriched
in b-quarks is selected from data by applying an impact parameter discriminant 
cut of 
$D>1$ on both event hemispheres and requiring the presence of  at least one
lepton with $p>3 \GeV,~p_t>1 \GeV$. The estimated b-purity of the 14929 
selected events is 97.5\%. 

Lepton hemispheres are classified in bins of ($1 \GeV \times 1 \GeV$) in the 
$(p,p_t)$ plane. The decay spectra in the context of the ACCMM model are 
generated as a function of $\mc$ and $\pf$,
and the MC $(p,p_t)$ distributions are modified accordingly. 
A binned likelihood function ${\cal L}$ is defined:
\begin{eqnarray}
   {\cal L}(\mc,\pf) = \prod_{i=1}^{\rm nbins} 
            \frac{{N^{\rm MC}_i(\mc,\pf)}^{N^{\rm DATA}_i}}{N^{\rm DATA}_i!}
            e^{-N^{\rm MC}_i(\mc,\pf)},
\end{eqnarray}

\noindent
where $N^{\rm DATA}_i$ and $N^{\rm MC}_i(\mc,\pf)$ are the number of events 
in the $i$th bin for data and MC, respectively. Only the shape is 
considered, that is, the total number of events in MC is
normalised in order to agree with the total number in data.
The data itself has no sensitivity for 
a simultaneous determination of $\mc$ and $\pf$. On the other hand, different 
values of $\mc$ and $\pf$ are found to reproduce the data with similar quality. 
Fixing the value of the c-quark mass to $\mc = 1.67 \GeV$ leads to a minimum
value of $\pf = 286 \pm 18 \MeV$, where the error is only statistical. Fixing 
the mass to $\mc = 1.60 \GeV$ moves the central value to $\pf = 346 \MeV$, but
leads to the same value of $\epsb^{\prime~{\rm REF}}$.
The quantity $-2 \log ({\cal L}/{\cal L_{\rm MC}})$, where 
${\cal L_{\rm MC}}$ is the value of the likelihood in MC at the minimum, 
should behave as a $\chi^2$ function. Its value is 1159.2 for 1274 degrees of 
freedom. The previous fit is repeated without a cut on $p_t$, with the result
$\pf = 273 \pm 17 \MeV$ and a $\chi^2$ of 1235.7 for 1340 degrees of freedom. 
A subsample with a harder lepton spectrum, $p>4 \GeV, p_t>1 \GeV$, 
gives $\pf = 288 \pm 20 \MeV$, with a similar fit quality. If no impact 
parameter discriminant cut is applied, the value obtained is still 
statistically consistent: $\pf = 272 \pm 13 \MeV$. 

  Several systematic effects are studied. A 1\% scale shift in the 
b-fragmentation parameter 
changes the value of $\pf$ by $14 \MeV$. A deterioration of the jet angular 
resolution by 20 mrad (half of the estimated resolution) leads to a shift of 
$20 \MeV$. Uncertainties in the momentum resolution give a smaller effect.
We estimate a total systematic uncertainty of $30 \MeV$, yielding:
\begin{eqnarray}
  \pf & = & 286 \pm 18\stat \pm 30\sys \MeV.
\end{eqnarray}

   Alternatively, the fit is performed in the context of the ISGW$^{**}$
model~\cite{isgw}. Using the D$^{**}$ fraction as a free parameter, we obtain:
\begin{eqnarray}
  f(\mathrm{D}^{**}) & = & (24 \pm 4\stat \pm 6\sys)\%,
\end{eqnarray}

\noindent
with a value $-2 \log ({\cal L}/{\cal L_{\rm MC}})=1150.3$ for 1274 degrees 
of freedom. The ISGW$^{**}$ model with this central value
leads to a slightly different value of 
$\epsb^{\prime~{\rm REF}}$. The difference is propagated as 
a systematic error due to the uncertainties in the modelling scheme. 

\begin{figure}[htbp]
\begin{center}
\includegraphics[width=\textwidth]{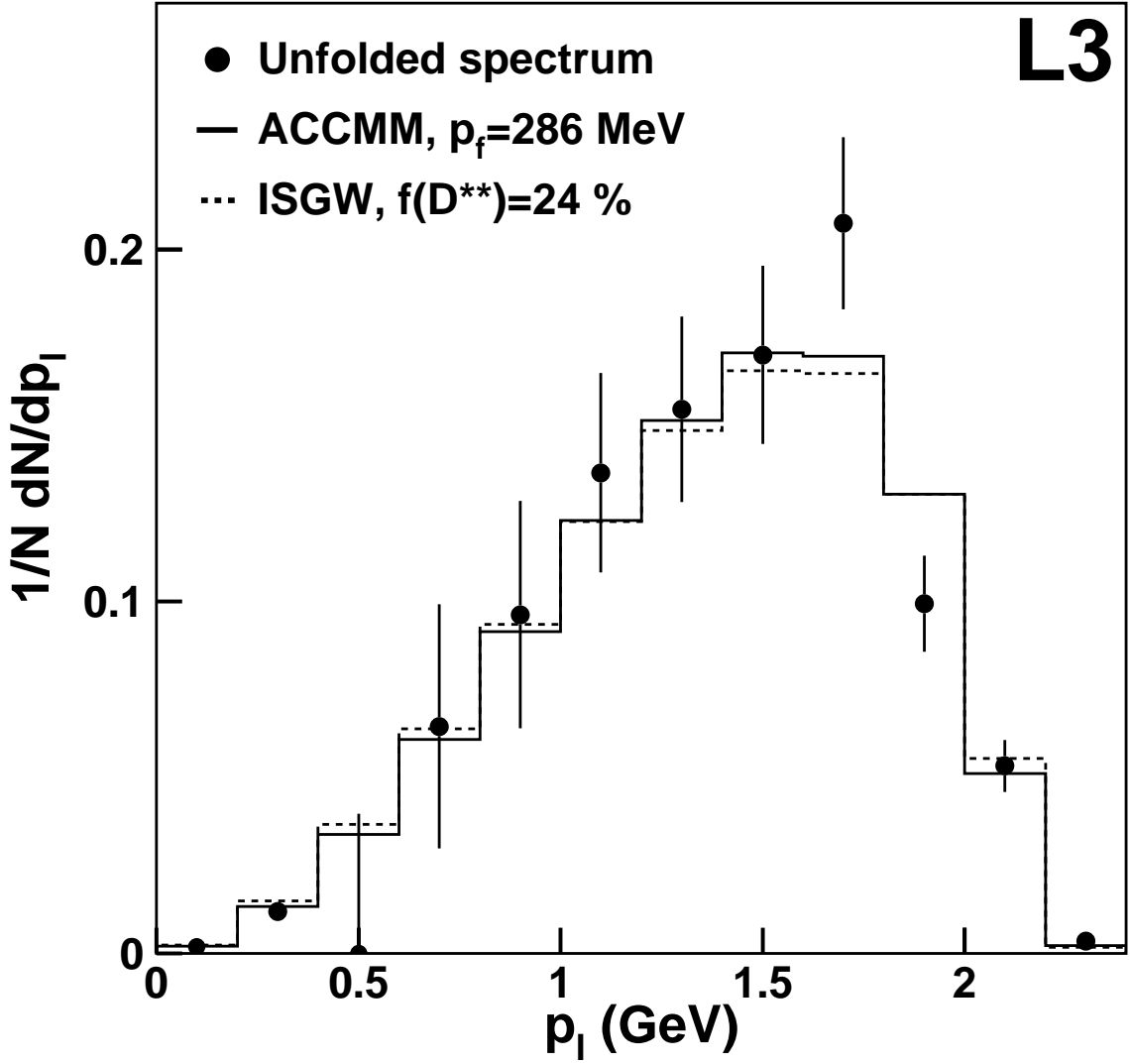}
\end{center}
\caption{\label{fig:unfold} Spectrum of the lepton momentum, $p_l$ 
in the centre-of-mass frame of the semileptonic decaying b-hadron.
The points and error bars are obtained by unfolding of the $(p,p_t)$ spectrum 
observed in data. All points are statistically correlated.
The histograms correspond to the optimal spectra 
favoured by the data in the context of ACCMM and ISGW$^{**}$ models. Overall 
consistency is observed.}
\end{figure}

   Finally, we perform a direct unfolding of the semileptonic decay momentum 
spectrum in the centre-of-mass frame of the b-hadron. For b-quarks, the original
MC spectrum in the range $0-2.4 \GeV$ is divided into 12 bins. 
The relative contents of these bins are adjusted in order to make 
the simulated and the observed $(p,p_t)$ spectra agree. The results are 
shown in Figure~\ref{fig:unfold}. There is good agreement with the optimal 
ACCMM, ISGW$^{**}$ spectra favoured by our data.

\subsection{Results}

  We use Equation \ref{eq:brl_formula} for the determination of $\Brbl$.
  Taking into account all the statistical and systematic errors on
$\epsb^{\prime}$ and $\epsb^{\prime~{\rm REF}}$, we obtain:

\begin{equation}
   \Brbl = (10.16 \pm 0.13\stat \pm 0.30\sys) \%.
\end{equation}

  Separate analyses, using only electron or muon tags, give consistent 
results within their statistical errors. 
The full breakdown of systematic errors is shown in Table~\ref{tab:blsys}.
 The total systematic error due to modelling uncertainties on $\Brbl$ is 
$0.2\%$. This error would increase to $0.3\%$ if the 
$f(\mathrm{D}^{**})$ range of $11\%$ to $32\%$ is used \cite{lephf}. 
However, this larger variation is excluded by our data at the $68\%$ 
confidence level.
The result represents an improvement on our previously published value 
\cite{l3btol92} and is consistent within errors, taking into 
account the different data samples, the different central values used in 
modelling and the fact that some systematic uncertainties are largely uncorrelated 
between the two measurements.

\begin{table}[htpb]
\begin{center}
\begin{tabular}{|l|r|}
\hline   
 Parameter & $\Delta \Brbl$ (\%) \\
\hline
\hline
 MC statistics & $+0.07$ \\
 $\mathrm{e}$ detector efficiency & $-0.07$ \\
 $\mu$ detector efficiency & $-0.11$ \\
 Decay parameters in $b \ra \ell$ & $+0.16$ \\
 Decay scheme in $b \ra \ell$ & $+0.11$ \\
 $<x_E({\rm b})>$ & $-0.06$ \\
 $\Brbcl$ & $+0.01$ \\
 $\Brbcbarl$ & $+0.02$ \\
 b-lifetime & $0.00$ \\
 $\Brbtaul$ & $+0.01$ \\
 $\Brbjl$ & $+0.01$ \\
\hline
 Uncertainty on $\Rc$ & 0.04 \\ 
 Uncertainty on $\epsc$ & 0.09 \\ 
 Uncertainty on $\epsuds$ & 0.10 \\ 
 Uncertainty in $\epsc^\prime$ ($\mathrm{c} \ra \ell$) & 0.06 \\ 
 Uncertainty in $\epsuds^\prime$ ($\ell$ in uds) & 0.01 \\ 
 Uncertainties on $\cb,\cb^\prime,\cb^{\prime\prime}$ & 0.08 \\ 
\hline
\hline
 Total & 0.30  \\
\hline
\end{tabular}
\end{center}
\caption{\label{tab:blsys} 
Systematic errors in the determination of $\Brbl$.
The dominant uncertainties are the 
detector efficiency, the semileptonic decay model and the uncertainties
coming from the use of the impact parameter tag in the determination 
of $\Rb$.
          The sign associated to a given
          error indicates the effect of a positive variation of the 
          corresponding parameter in Table \ref{tab:epsilonbl_mc}.}
\end{table}

The semileptonic branching ratio for mesons containing a b-quark is
extracted as follows:
\begin{eqnarray}
   \BrBl & = & \frac{\Brbl-f_{\Lb}~\BrLl} {1-f_{\Lb}},
\end{eqnarray}

\noindent 
where $f_{\Lb}$ is the fraction of b-baryons in Z decays
and $\BrLl$ is the semileptonic branching ratio in b-baryon decays. Assuming 
that all B-mesons have the same semileptonic branching ratio, our
result is:
\begin{eqnarray}
   \BrBl & = & (10.47 \pm 0.36\stasys~~^{+0.13}_{-0.10}~(f_{\Lb})~~^{+0.12}_{-0.13}~(\BrLl))\%
\end{eqnarray}

\noindent 
where the last two errors are due to the uncertainties on
the present measurements of 
$f_{\Lb} = (10.1^{+3.9}_{-3.1})\%$ \cite{pdg98}
and $\BrLl = (7.4 \pm 1.1)\%$ \cite{lambdab_semilept}. A similar conclusion is
reached if the experimental values of the inclusive ($\taub$) and 
b-baryon ($\tau_{\Lb}$) lifetimes \cite{pdg98} are used together with the 
assumption: $\BrLl/\tau_{\Lb}=\Brbl/\taub$.

\section {Conclusion}

  Using double-tag methods on one million hadronic Z events collected in 
1994 and 1995, we 
determine the values of $\Rb = \Gamma(\mathrm{Z \rightarrow b\bar{b}}) / 
\Gamma(\mathrm{Z} \rightarrow \mbox{hadrons})$ 
and of the semileptonic branching ratio of
b-quarks, $\Brbl$. Two methods of tagging the presence of b-quarks are used.
The first method
exploits the capabilities of the silicon microvertex detector and is based
on the large impact parameter of tracks from weak b-decays with
respect to the $\ee$ collision point. In the second method, a
high-$\pt$ lepton tag is used to enrich the b-content of the sample and its 
$(p,\pt)$ spectrum is used to constrain the model dependent uncertainties 
in the semileptonic b-decay. We measure:
\begin{eqnarray}
    \Rb & = & 0.2174 \pm 0.0015\stat \pm 0.0028\sys, \\
    \Brbl & = & (10.16 \pm 0.13\stat \pm 0.30\sys)\%.
\end{eqnarray}
 
   The measurement of $\Rb$ agrees with the Standard Model expectations
and with previous determinations \cite{otherlep}. The measurement of 
$\Brbl$ is in agreement with measurements of the semileptonic branching 
ratio of B mesons at the $\Upsilon(4S)$ \cite{4S_semilep}.
  
%
%
\section*{Acknowledgments}

We wish to
express our gratitude to the CERN accelerator divisions for
the excellent performance of the LEP machine.
We acknowledge the contributions of the engineers
and technicians who have participated in the construction
and maintenance of this experiment.

\bibliographystyle{l3style}

\newpage
\input namelist180.tex

\newpage

\end{document}

%% file: namelist180.tex
\typeout{   }     
\typeout{Using author list for paper 180 -?}
\typeout{$Modified: Wed Aug 25 10:07:39 1999 by clare $}
\typeout{!!!!  This should only be used with document option a4p!!!!}
\typeout{   }
%
%
%
%
%
%

\newcount\tutecount  \tutecount=0
\def\tutenum#1{\global\advance\tutecount by 1 \xdef#1{\the\tutecount}}
\def\tute#1{$^{#1}$}
\tutenum\aachen            
\tutenum\nikhef            
\tutenum\mich              
\tutenum\lapp              
\tutenum\basel             
\tutenum\lsu               
\tutenum\beijing           
\tutenum\berlin            
\tutenum\bologna           
\tutenum\tata              
\tutenum\ne                
\tutenum\bucharest         
\tutenum\budapest          
\tutenum\mit               
\tutenum\debrecen          
\tutenum\florence          
\tutenum\cern              
\tutenum\wl                
\tutenum\geneva            
\tutenum\hefei             
\tutenum\seft              
\tutenum\lausanne          
\tutenum\lecce             
\tutenum\lyon              
\tutenum\madrid            
\tutenum\milan             
\tutenum\moscow            
\tutenum\naples            
\tutenum\cyprus            
\tutenum\nymegen           
\tutenum\caltech           
\tutenum\perugia           
\tutenum\cmu               
\tutenum\prince            
\tutenum\rome              
\tutenum\peters            
\tutenum\salerno           
\tutenum\ucsd              
\tutenum\santiago          
\tutenum\sofia             
\tutenum\korea             
\tutenum\alabama           
\tutenum\utrecht           
\tutenum\purdue            
\tutenum\psinst            
\tutenum\zeuthen           
\tutenum\eth               
\tutenum\hamburg           
\tutenum\taiwan            
\tutenum\tsinghua          
{
\parskip=0pt
\noindent
{\bf The L3 Collaboration:}
\ifx\selectfont\undefined
 \baselineskip=10.8pt
 \baselineskip\baselinestretch\baselineskip
 \normalbaselineskip\baselineskip
 \ixpt
\else
 \fontsize{9}{10.8pt}\selectfont
\fi
\medskip
\tolerance=10000
\hbadness=5000
\raggedright
\hsize=162truemm\hoffset=0mm
\def\r{\rlap,}
\noindent

M.Acciarri\r\tute\milan\
P.Achard\r\tute\geneva\ 
O.Adriani\r\tute{\florence}\ 
M.Aguilar-Benitez\r\tute\madrid\ 
J.Alcaraz\r\tute\madrid\ 
G.Alemanni\r\tute\lausanne\
J.Allaby\r\tute\cern\
A.Aloisio\r\tute\naples\ 
M.G.Alviggi\r\tute\naples\
G.Ambrosi\r\tute\geneva\
% ...
\newpage
\begin{list}{A}{\itemsep=0pt plus 0pt minus 0pt\parsep=0pt plus 0pt minus 0pt
                \topsep=0pt plus 0pt minus 0pt}
\item[\aachen]
 I. Physikalisches Institut, RWTH, D-52056 Aachen, FRG$^{\S}$\\
 III. Physikalisches Institut, RWTH, D-52056 Aachen, FRG$^{\S}$
\item[\nikhef] National Institute for High Energy Physics, NIKHEF, 
     and University of Amsterdam, NL-1009 DB Amsterdam, The Netherlands
\item[\mich] University of Michigan, Ann Arbor, MI 48109, USA
\item[\lapp] Laboratoire d'Annecy-le-Vieux de Physique des Particules, 
     LAPP,IN2P3-CNRS, BP 110, F-74941 Annecy-le-Vieux CEDEX, France
\item[\basel] Institute of Physics, University of Basel, CH-4056 Basel,
     Switzerland
\item[\lsu] Louisiana State University, Baton Rouge, LA 70803, USA
\item[\beijing] Institute of High Energy Physics, IHEP, 
  100039 Beijing, China$^{\triangle}$ 
\item[\berlin] Humboldt University, D-10099 Berlin, FRG$^{\S}$
\item[\bologna] University of Bologna and INFN-Sezione di Bologna, 
     I-40126 Bologna, Italy
\item[\tata] Tata Institute of Fundamental Research, Bombay 400 005, India
\item[\ne] Northeastern University, Boston, MA 02115, USA
\item[\bucharest] Institute of Atomic Physics and University of Bucharest,
     R-76900 Bucharest, Romania
\item[\budapest] Central Research Institute for Physics of the 
     Hungarian Academy of Sciences, H-1525 Budapest 114, Hungary$^{\ddag}$
\item[\mit] Massachusetts Institute of Technology, Cambridge, MA 02139, USA
\item[\debrecen] Lajos Kossuth University-ATOMKI, H-4010 Debrecen, Hungary$^\P$
\item[\florence] INFN Sezione di Firenze and University of Florence, 
     I-50125 Florence, Italy
\item[\cern] European Laboratory for Particle Physics, CERN, 
     CH-1211 Geneva 23, Switzerland
\item[\wl] World Laboratory, FBLJA  Project, CH-1211 Geneva 23, Switzerland
\item[\geneva] University of Geneva, CH-1211 Geneva 4, Switzerland
\item[\hefei] Chinese University of Science and Technology, USTC,
      Hefei, Anhui 230 029, China$^{\triangle}$
\item[\seft] SEFT, Research Institute for High Energy Physics, P.O. Box 9,
      SF-00014 Helsinki, Finland
\item[\lausanne] University of Lausanne, CH-1015 Lausanne, Switzerland
\item[\lecce] INFN-Sezione di Lecce and Universit\'a Degli Studi di Lecce,
     I-73100 Lecce, Italy
\item[\lyon] Institut de Physique Nucl\'eaire de Lyon, 
     IN2P3-CNRS,Universit\'e Claude Bernard, 
     F-69622 Villeurbanne, France
\item[\madrid] Centro de Investigaciones Energ{\'e}ticas, 
     Medioambientales y Tecnolog{\'\i}cas, CIEMAT, E-28040 Madrid,
     Spain${\flat}$ 
\item[\milan] INFN-Sezione di Milano, I-20133 Milan, Italy
\item[\moscow] Institute of Theoretical and Experimental Physics, ITEP, 
     Moscow, Russia
\item[\naples] INFN-Sezione di Napoli and University of Naples, 
     I-80125 Naples, Italy
\item[\cyprus] Department of Natural Sciences, University of Cyprus,
     Nicosia, Cyprus
\item[\nymegen] University of Nijmegen and NIKHEF, 
     NL-6525 ED Nijmegen, The Netherlands
\item[\caltech] California Institute of Technology, Pasadena, CA 91125, USA
\item[\perugia] INFN-Sezione di Perugia and Universit\'a Degli 
     Studi di Perugia, I-06100 Perugia, Italy   
\item[\cmu] Carnegie Mellon University, Pittsburgh, PA 15213, USA
\item[\prince] Princeton University, Princeton, NJ 08544, USA
\item[\rome] INFN-Sezione di Roma and University of Rome, ``La Sapienza",
     I-00185 Rome, Italy
\item[\peters] Nuclear Physics Institute, St. Petersburg, Russia
\item[\salerno] University and INFN, Salerno, I-84100 Salerno, Italy
\item[\ucsd] University of California, San Diego, CA 92093, USA
\item[\santiago] Dept. de Fisica de Particulas Elementales, Univ. de Santiago,
     E-15706 Santiago de Compostela, Spain
\item[\sofia] Bulgarian Academy of Sciences, Central Lab.~of 
     Mechatronics and Instrumentation, BU-1113 Sofia, Bulgaria
\item[\korea] Center for High Energy Physics, Adv.~Inst.~of Sciences
     and Technology, 305-701 Taejon,~Republic~of~{Korea}
\item[\alabama] University of Alabama, Tuscaloosa, AL 35486, USA
\item[\utrecht] Utrecht University and NIKHEF, NL-3584 CB Utrecht, 
     The Netherlands
\item[\purdue] Purdue University, West Lafayette, IN 47907, USA
\item[\psinst] Paul Scherrer Institut, PSI, CH-5232 Villigen, Switzerland
\item[\zeuthen] DESY, D-15738 Zeuthen, 
     FRG
\item[\eth] Eidgen\"ossische Technische Hochschule, ETH Z\"urich,
     CH-8093 Z\"urich, Switzerland
\item[\hamburg] University of Hamburg, D-22761 Hamburg, FRG
\item[\taiwan] National Central University, Chung-Li, Taiwan, China
\item[\tsinghua] Department of Physics, National Tsing Hua University,
      Taiwan, China
\item[\S]  Supported by the German Bundesministerium 
        f\"ur Bildung, Wissenschaft, Forschung und Technologie
\item[\ddag] Supported by the Hungarian OTKA fund under contract
numbers T019181, F023259 and T024011.
\item[\P] Also supported by the Hungarian OTKA fund under contract
  numbers T22238 and T026178.
\item[$\flat$] Supported also by the Comisi\'on Interministerial de Ciencia y 
        Tecnolog{\'\i}a.
\item[$\sharp$] Also supported by CONICET and Universidad Nacional de La Plata,
        CC 67, 1900 La Plata, Argentina.
\item[$\diamondsuit$] Also supported by Panjab University, Chandigarh-160014, 
        India.
\item[$\triangle$] Supported by the National Natural Science
  Foundation of China.
\item[\dag] Deceased.
\end{list}
}
\vfill



